\documentclass[aps,12pt,a4paper,prd,showpacs,superscriptaddress,floatfix]{revtex4-2}
\usepackage[T1]{fontenc}
\usepackage[latin9]{inputenc}
\usepackage{amsmath}
\usepackage{graphicx}
\usepackage{babel,comment}
\usepackage{color,diagbox}
\usepackage{subfigure}
\graphicspath{{pic/}}

\begin{document}
\title{Pion Condensation and Pion Star from Holographic QCD}

\author{Yidian Chen}
\email{chenyidian@hznu.edu.cn}
\affiliation{School of Physics, Hangzhou Normal University, Hangzhou, 311121, P.R. China}

\author{Mingshan Ding}
\email{13149710203@163.com}
\affiliation{School of Physics, Hangzhou Normal University, Hangzhou, 311121, P.R. China}

\author{Danning Li}
\email[]{lidanning@jnu.edu.cn}
\affiliation{Department of Physics and Siyuan Laboratory, Jinan University, Guangzhou 510632, P.R. China}

\author{Kazem Bitaghsir Fadafan}
\email[]{bitaghsir@shahroodut.ac.ir}
\affiliation{Faculty of Physics, Shahrood University of Technology, P.O.Box 3619995161 Shahrood, Iran}

\author{Mei Huang}
\email[]{huangmei@ucas.ac.cn}
\affiliation{School of Nuclear Science and Technology, University of Chinese Academy of Sciences, Beijing 100049, China}

\begin{abstract}
The properties of QCD matter at finite isospin densities are investigated employing holographic hard-wall and soft-wall AdS/QCD models. It is confirmed that at high enough isospin densities, charged pions start to condense and the pion superfluid phase appears in the system. It is shown that the chiral condensate and the pion condensate can be transformed to each other and form a `chiral circle' in the superfluid phase. We derived the Equation of State (EoS) for pionic matter, calculated the normalized trace anomaly $\Delta$ and $(\epsilon-3p)/m_\pi^4$, and analyzed the sound speed and adiabatic index. Additionally, we provided data on the mass-radius relation and tidal deformability of pion stars. The results indicate that the holographic models align well with lattice QCD concerning isospin density, axial-vector condensation, EoS, and trace anomaly, though discrepancies in sound speed and adiabatic index emerge at higher isospin chemical potentials. The holographic models closely match those from chiral perturbation theory ($\chi$PT), suggesting that they can be considered as five-dimensional description of $\chi$PT.
\end{abstract}
\maketitle

\section{Introduction}

Under the condition of finite isospin chemical potential, Quantum Chromodynamics (QCD) exhibits a phenomenon: the condensation of charged pions. This phenomenon can occur in various physical environments, including heavy-ion collision experiments, the interiors of neutron stars, and the very early universe \cite{Schwarz:2009ii}. In the core of a neutron star, the excess of down quarks over up quarks generates an isospin chemical potential $\mu_I$ \cite{Migdal:1990vm}. Although the potential is much smaller than the baryon chemical potential $\mu_I\ll\mu_B$, it may explain some very heavy neutron stars \cite{Lattimer:2006xb}. Furthermore, if the early universe possessed a significant lepton flavor asymmetry, this asymmetry could been converted into a larger isospin chemical potential as the universe expanded, triggering pion condensation. This condensation could influence not only the evolution of the early universe but also leave a distinctive imprint on the spectrum of primordial gravitational waves and the mass distribution of primordial black holes \cite{Liebling:2012fv,Vovchenko:2020crk}, offering valuable clues for studying the early universe.

Pion condensation in the early universe may also lead to another phenomenon: the formation of pion stars \cite{Vartanyan:1984zz,Brandt:2018bwq,Carignano:2016lxe}. As one type of bosonic stars\cite{Jetzer:1991jr}, pion stars are formed due to pion condensation when the pion density in space region becomes high enough, thus can gravitationally attract each other to form a self-gravitating configuration overcoming their quantum-mechanical repulsion \cite{Kleihaus:2009kr}. These hypothetical stars could reveal their existence through neutrino and photon signals released during their evaporation process, or through the observation of gravitational waves in binary systems. Although the stability and lifetime of pion stars remain an open question, their lifetime is relatively short, and they could impact Big Bang nucleosynthesis, potentially affecting the primordial abundance of nuclei.

Due to the sign problem in lattice QCD \cite{Barbour:1986jf,Kogut:1994eq}, direct calculations at finite baryon chemical potential are infeasible. Fortunately, lattice QCD calculations can be performed at finite isospin chemical potential without encountering the sign problem \cite{Kogut:2002tm}. Significant progress has been made in lattice QCD simulations at finite isospin chemical potential, including confirming the second-order phase transition of Bose-Einstein condensation (BEC) of charged pions \cite{Kogut:2002zg,Kogut:2004zg,deForcrand:2007uz,Cea:2012ev,Detmold:2012wc,Endrodi:2014lja,Brandt:2017oyy,Brandt:2018omg}, studying pion dynamics, the pionic equation of state \cite{Brandt:2018bwq}, and sound speed at low temperatures \cite{Brandt:2022hwy}. Beyond lattice QCD, various models like $\chi$PT \cite{Kaplan:1986yq,Son:2000xc,Carignano:2016lxe}, Nambu-Jona-Lasinio \cite{Toublan:2003tt,Barducci:2004tt,He:2005nk,Sun:2007fc,Xia:2013caa,Chao:2018ejd,Zhang:2018ome}, quark-meson \cite{Adhikari:2018cea,Wang:2017vis}, linear sigma \cite{Loewe:2013coa,Wang:2015bky}, random matrix \cite{Klein:2003fy}, and perturbative QCD \cite{Graf:2015pyl} have been exploring QCD phase transitions and pionic matter at finite isospin chemical potential.

Inspired by the AdS/CFT correspondence principle \cite{Maldacena:1997re,Gubser:1998bc,Witten:1998qj}, the bottom-up holographic QCD (HQCD) approach has become a powerful tool for studying strongly coupled gauge theories \cite{Gursoy:2007cb,Gursoy:2007er,Gubser:2008ny,Gubser:2008yx,DeWolfe:2010he,Li:2013oda,Li:2014hja,Li:2014dsa,BitaghsirFadafan:2018uzs,Chen:2019rez,Chen:2022goa,Cai:2022omk}. This method constructs an effective field theory in a higher-dimensional gravitational framework to capture the main features of QCD. Recent studies have used holographic models to explore QCD matter at finite isospin chemical potential. The properties of pionic matter have been studied in the Witten-Sakai-Sugimoto model \cite{Kovensky:2023mye,Kovensky:2024oqx}, also pion condensation \cite{Nishihara:2014nsa,Kim:2007xi,Parnachev:2007bc,Aharony:2007uu,Basu:2008bh,Ammon:2009fe,Lee:2013oya,Albrecht:2010eg,Lv:2018wfq}, the QCD phase diagram \cite{Nishihara:2014nva,Cao:2020ske}, and pion dynamics \cite{Cao:2022csq,Liang:2023lgs}. The application of HQCD to neutron stars offers new insights into their extreme conditions \cite{Jokela:2018ers,Jarvinen:2021jbd,Zhang:2022uin}. These models align with astrophysical observations and have broader applications, including superconductivity \cite{BitaghsirFadafan:2018iqr,BitaghsirFadafan:2019ofb,BitaghsirFadafan:2020otb}. In this paper, we extend such application to study pionic stars.

HQCD approach provides a framework to link the dynamics of an infinite number of QCD hadrons with low-energy effective theories, such as $\chi$PT. By integrating out heavy Kaluza-Klein modes and keeping only the lowest-lying mesons, HQCD can be reformulated into chiral effective Lagrangians \cite{Colangelo:2012ipa}. This approach results in all low-energy constants of the $\chi$PT being expressed in terms of holographic integrals, thereby directly connecting the holographic models with chiral effective theories used in QCD. Notably, the Gell-Mann-Oakes-Renner relation, which relates the pion mass to the quark condensate and quark masses, is naturally reproduced at the lowest order of the derivative expansion in this framework \cite{Harada:2014lza}.

The structure of this paper is as follows: Sec. \ref{sec:model} introduces the five-dimensional holographic model used in our study. Sec. \ref{sec:eos} details the calculation of the pion state equation. Sec. \ref{sec:cs} presents the calculation of the sound speed and the adiabatic index. Sec. \ref{sec:star} discusses the mass-radius relation and tidal deformability of pion stars. Finally, Sec. \ref{sec:sum} summarizes and discusses our findings.

\section{Model setup}
\label{sec:model}

The (improved) soft-wall (SW) and hard-wall (HW) models provide a bottom-up holographic framework to explore QCD properties including hadron spectra \cite{Erlich:2005qh,DaRold:2005vr,Karch:2006pv}, chiral phase transitions \cite{Chelabi:2015cwn,Chelabi:2015gpc}, and pion condensation \cite{Nishihara:2014nsa,Nishihara:2014nva,Cao:2020ske}. In the models, the real part of the complex scalar field $X$ represents sigma condensation, while its imaginary part describes the phase of the pion condensate under nonzero isospin chemical potential. These models are constructed using the probe approximation within the AdS$_5$ or AdS-Schwarzschild black hole spacetime.

The background metric is given as
\begin{eqnarray}
    ds^2=\frac{L^2}{z^2}[-f(z)dt^2+d\vec{x}^2+\frac{dz^2}{f(z)}],
\end{eqnarray}
with AdS radius $L=1$, blackening factor $f(z)=1-\frac{z^4}{z_h^4}$ and the horizon $z_h$.

The holographic QCD model with $N_f=2$ is constructed from $SU(2)_L\times SU(2)_R\equiv SU(2)_V\times SU(2)_A$ flavor symmetry. The Lagrangian density can be written as
\begin{eqnarray}
    \mathcal{L}_5={\rm Tr}[|D_M X|^2+m_5^2(z)|X|^2+\lambda|X|^4+\frac{1}{4g_5^2}(F_{L,MN}^2+F_{R,MN}^2)],
\end{eqnarray}
with the coupling $g_5^2=12\pi^2/N_c=4\pi^2$ \cite{Erlich:2005qh}. The scalar field $X$ and gauge fields $A_{L/R,M}$ are dual to the quark condensate and current operators, respectively. The $m_5$ is 5-dimensional mass and $\lambda$ is quartic coupling. The vacuum of scalar can be given as
\begin{eqnarray}
    X=\frac{1}{2}(\Sigma\sigma^0+i\Pi^a\sigma^a),
\end{eqnarray}
with Pauli matrices $\sigma^a$ and identity matrix $\sigma^0$. Here, $\Sigma$ and $\Pi$ corresponds to sigma condensate $\langle \bar{q}q\rangle$ and pion condensate $\langle \bar{q}\gamma^5q\rangle$, respectively. The covariant derivative $D_MX$ and gauge field strength $F_{L/R}^{MN}$ are written as
\begin{eqnarray}
    D_MX&=&\partial_MX-iA_{L,M}X+iXA_{R,M},\\
    F_{L/R}^{MN}&=&\partial^MA_{L/R}^N-\partial^NA_{L/R}^M-i[A_{L/R}^M,A_{L/R}^N],
\end{eqnarray}
with $A_{L/R,M}=A_{L/R,M}^at^a$ and generators $t^a$ of $SU(2)$ group. The chiral gauge field can be redefined through vector and axial-vector fields,
\begin{eqnarray}
    A_{L,M}=V_M+A_M,\qquad A_{R,M}=V_M-A_M,
\end{eqnarray}
then the covariant derivative and gauge field strength are rewritten as
\begin{eqnarray}
    D_MX&=&\partial_MX-i[V_M,X]-i\{A_M,X\},\\
    F_V^{MN}&=&\partial^MV^N-\partial^NV^M-i[V^M,V^N]-i[A^M,A^N],\\
    F_A^{MN}&=&\partial^MA^N-\partial^NA^M-i[V^M,A^N]-i[A^M,V^N].
\end{eqnarray}
In the following, the gauge $A_{L/R}^z=0$ is considered. 

As discussed in Refs. \cite{Nishihara:2014nsa,Nishihara:2014nva,Cao:2020ske}, the non-vanishing condensate can be chosen as $\Pi^{(1)}\equiv \Pi$ under $U(1)_I$ symmetry. Moreover, the nonzero fields of model are $A_0^{(2)}\equiv a_2$, $V_0^{(3)}\equiv v$, and $\Sigma$. From the holographic dictionary, the gauge fields can be expanded at the AdS boundary ($z\to 0$) as
\begin{eqnarray}
    v\rightarrow \mu_I-C_1 n_I z^2+\mathcal{O}(z^2),\qquad a_2\rightarrow C_2\langle\sigma_A\rangle z^2+\mathcal{O}(z^2),
\end{eqnarray}
with isospin chemical potential $\mu_I$, isospin density $n_I$, axial-vector condensate $\langle\sigma_A\rangle$, and normalized constants $C_{1,2}$. In this paper, isospin density and axial-vector condensate are obtained by (\ref{eq:nI},\ref{eq:sigmaA}), so the exact values of constants $C_{1,2}$ are not important.
For the (pseudo-)scalar fields, the expansions are
\begin{eqnarray}\label{eq:uv-exp}
    \Sigma\rightarrow m_q \zeta z+\frac{\langle\sigma\rangle}{\zeta}z^3+\mathcal{O}(z^3),\qquad \Pi^a\rightarrow \frac{\langle\pi^a\rangle}{\zeta}z^3+\mathcal{O}(z^3),
\end{eqnarray}
with normalization constant $\zeta$, quark mass $m_q$, sigma condensate $\langle\sigma\rangle$, and pion condensate $\langle\pi^a\rangle$.

\subsection{Hard-Wall Model}

The zero temperature is considered in the hard-wall model. So, the horizon $z_h$ tends to infinity $z_h\to\infty$, and the blacken factor is $f(z)=1$. The full action can be written as \cite{Nishihara:2014nsa,Nishihara:2014nva}
\begin{eqnarray}
    S_5 &=& -\int d^4x \int_\epsilon^{z_m} dz \sqrt{-g}(\mathcal{L}_5+\mathcal{L}_{BD}),\\
    \mathcal{L}_{BD} &=& {\rm Tr}\{\lambda_4 z_m|X|^4-m_2^2z_m|X|^2\}\delta(z-z_m),
\end{eqnarray}
with ultra-violate (UV) and infra-red (IR) cut-off $\epsilon$ and $z_m$. The UV cut-off can be viewed as a small number that tends to zero. The Ref. \cite{DaRold:2005vr} introduces the no-dimensional parameters $\lambda_4$ and $m_2$ at the boundary. The 5D mass square $m_5^2(z)$ is viewed as a constant in the hard-wall model. According to AdS/CFT correspondence, the 5D mass is $m_5^2=\Delta(\Delta-4)=-3$ with conformal dimension $\Delta=3$. The values of the parameters are chosen as Refs. \cite{Nishihara:2014nsa,Nishihara:2014nva} and shown in Tab. \ref{tab:paras}. The equations of motion are given as
\begin{eqnarray}\label{eq:hw-1}
    a_2\Pi v-\left(\frac{3}{z^2}+a_2^2\right)\Sigma+\frac{3}{z}\Sigma'-\Sigma''&=&0,\\
    v\Sigma a_2-\left(\frac{3}{z^2}+v^2\right)\Pi+\frac{3}{z}\Pi'-\Pi''&=&0,\\
    \frac{g_5^2\Sigma}{z^2}\left(v\Pi -a_2\Sigma\right)-\frac{a_2'}{z}+a_2''&=&0,\\
    \frac{g_5^2\Pi}{z^2}\left(a_2\Sigma -v\Pi\right)-\frac{v'}{z}+v''&=&0. \label{eq:hw-4}
\end{eqnarray}
The following boundary conditions are considered for solving the coupled second-order differential equations. At the UV ($z\to\epsilon$), the boundary conditions are
\begin{eqnarray}\label{eq:hw-bc-uv}
    \frac{\Sigma}{z}\bigg|_{\epsilon}=m_q\zeta,\qquad v\big|_{\epsilon}=\mu_I,\qquad \frac{\Pi}{z}\big|_{\epsilon}=0,\qquad a_2\big|_{\epsilon}=0.
\end{eqnarray}
with normalization constant $\zeta=1$. At the IR ($z\to z_m$), the boundary conditions are
\begin{equation}
    \begin{aligned}
        \partial_z\Sigma\big|_{z_m}&=-\frac{\Sigma}{2z_m}\bigg(\lambda_4(\Sigma^2+\pi^2)-2m_2^2\bigg)\bigg|_{z_m},\qquad \partial_z v\big|_{z_m}&\!\!\!\!\!=0,\\
        \partial_z\Pi\big|_{z_m}&=-\frac{\Pi}{2z_m}\bigg(\lambda_4(\Sigma^2+\pi^2)-2m_2^2\bigg)\bigg|_{z_m},\qquad \partial_z a_2\big|_{z_m}&\!\!\!\!\!=0.
    \end{aligned}
\end{equation}

\begin{figure}[tbp]
    \centering
    \subfigure[]{\label{subfig:condensate}
    \includegraphics[width=0.4\textwidth,height=0.3\textwidth]{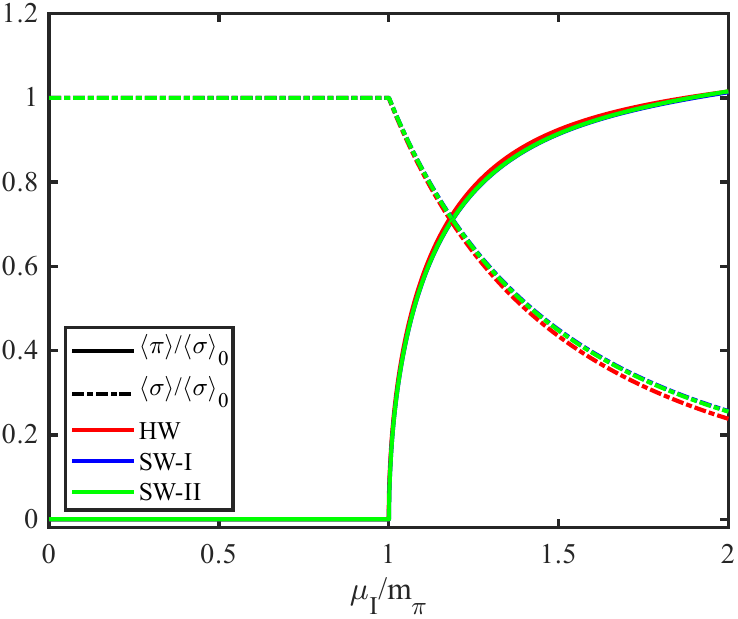}}
    \hspace{0.01\linewidth}
    \subfigure[]{\label{subfig:circle}
    \includegraphics[width=0.4\textwidth,height=0.3\textwidth]{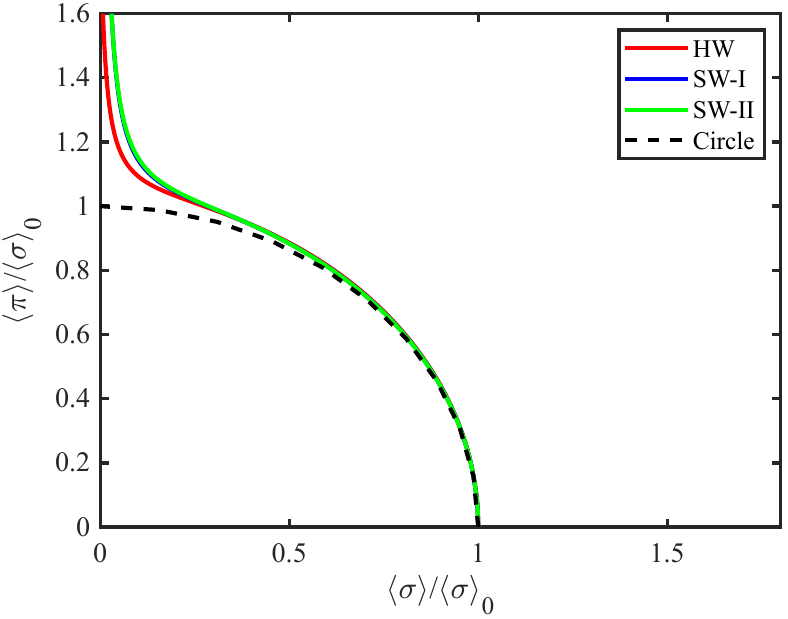}}
    \caption{\label{fig:condensate} Panel (a) illustrates how the sigma and pion condensates vary with the isospin chemical potential $\mu_I$ in both the HW and SW models. Panel (b) reveals the interrelationship between the sigma and pion condensates, where the black dashed line represents the unit circle.}
\end{figure}

By applying UV and IR boundary conditions, combined with the Chebyshev spectral method, the equations of motion (\ref{eq:hw-1}-\ref{eq:hw-4}) can be solved. Furthermore, through UV expansion \eqref{eq:uv-exp}, we can obtain the condensations of sigma and pion. The condensates as a function of isospin chemical potential $\mu_I$ are shown in Fig. \ref{fig:condensate}.

\begin{table*}[tbp]
    \centering
    \tabcolsep=0.5em
    \begin{tabular}{c|c c | c c c | c c c c}
	\hline
	Parameters & $m_q$ (MeV) & $\lambda$ & $z_m^{-1}$ (MeV) & $\lambda_4$ & $m_2^2$ & $\mu_g$ (MeV) & $\mu_c$ (MeV) & $\gamma_m$ & $\kappa$ \\
	\hline
	HW & 2.29 & 0 & 323 & 4.4 & 5.39 & - & - & - & - \\
	SW-I & 3.22 & 80 & - & - & - & 440 & 1450 & - & - \\
	SW-II & 3.58 & 14.7 & - & - & - & 440 & - & 3.7 & 1 \\
	\hline
    \end{tabular}
    \caption{\label{tab:paras} The parameters of the hard-wall (HW) model and soft-wall (SW) model, taken from \cite{Nishihara:2014nsa,Nishihara:2014nva,Cao:2020ske,Liang:2023lgs}.}
\end{table*}

\subsection{Soft-Wall Model}

In the soft wall model, we have set the temperature values at 2 MeV, 5 MeV, 10 MeV, and 15 MeV. Numerical calculations show negligible differences in the results at different temperatures. Therefore, we have chosen the 10 MeV (with horizon $z_h=1/\pi T\simeq 31.8$ GeV$^{-1}$) temperature for presentation. This temperature is small compared to the pion mass $m_\pi\simeq 140$ MeV and can be approximated as zero temperature. The full action is given as
\begin{eqnarray}
    S_5 &=& -\int d^4x \int_\epsilon^{z_h} dz \sqrt{-g}e^{-\Phi(z)}\mathcal{L}_5,
\end{eqnarray}
where $\Phi(z)=\mu_g^2z^2$ is dilaton field with constant $\mu_g$ that match the Regge slope of light mesons. To describe the mesons spectra and chiral condensate simultaneously, we have considered two possible forms of the five-dimensional mass squared $m_5^2(z)$ that depend on the fifth dimension, as follows:
\begin{subequations}
\begin{align}
    m_5^2(z)&=-3-\mu_c^2z^2, &(\text{SW-I})\\
    m_5^2(z)&=-3\bigg(1+\gamma_m\tanh{[\kappa\Phi(z)]}\bigg), & (\text{SW-II})
\end{align}
\end{subequations}
with free parameter $\mu_c$, $\gamma_m$, and $\kappa$. The leading term of them are consistent with the hard-wall model $m_5^2=-3$. The values of the parameters are given from Refs. \cite{Cao:2020ske,Liang:2023lgs} and shown in Tab. \ref{tab:paras}. The equations of motion of the soft-wall model are obtained as
\begin{eqnarray}
    \frac{\lambda\Sigma^3}{2z^2}+\frac{v\Pi a_2}{f}+\left(-\frac{a_2^2}{f}+\frac{\lambda\Pi^2+2m_5^2(z)}{2z^2}\right)\Sigma+\left(-f'+f \left(\frac{3}{z}+\Phi'\right)\right)\Sigma'-f\Sigma''&=&0,\\
    \frac{\lambda\Pi ^3}{2 z^2}+\frac{v\Sigma a_2}{f}+\left(-\frac{v^2}{f}+\frac{\lambda\Sigma ^2+2 m_5^2(z)}{2 z^2}\right)\Pi +\left(-f'+f \left(\frac{3}{z}+\Phi '\right)\right)\Pi '-f \Pi ''&=&0,\\
    \frac{g_5^2\Sigma}{z^2}\left(v\Pi -a_2\Sigma\right)+\left(-\frac{f}{z}-f\Phi'\right)a_2'+fa_2''&=&0,\\
    \frac{g_5^2\Pi}{z^2}\left(a_2\Sigma-v\Pi\right)+\left(-\frac{f}{z}-f\Phi'\right)v'+fv''&=&0.
\end{eqnarray}
At the UV, the boundary conditions are the same as Eq. \eqref{eq:hw-bc-uv}. At the IR ($z\to z_h$), the expansions of $\Sigma$, $\Pi$, $a_2$, and $v$ have the following form
\begin{equation}
    \begin{aligned}
        \Sigma(z)&\rightarrow \Sigma_0+\frac{6\Sigma_0+2z_h^2\mu_c^2\Sigma_0-\lambda\Pi_0^2\Sigma_0-\lambda\Sigma_0^3}{8z_h}(z-z_h)+\mathcal{O}(z-z_h),\\
        \Pi(z)&\rightarrow \Pi_0+\frac{6\Pi_0+2z_h^2\mu_c^2\Pi_0-\lambda\Sigma_0^2\Pi_0-\lambda\Pi_0^3}{8z_h}(z-z_h)+\mathcal{O}(z-z_h),\\
        a_2(z)&\rightarrow a_{2,1}(z-z_h)+\mathcal{O}(z-z_h),\\
        v(z)&\rightarrow v_{1}(z-z_h)+\mathcal{O}(z-z_h),\\
    \end{aligned}
\end{equation}
with integral constants $\Sigma_0$, $\Pi_0$, $a_{2,1}$, and $v_1$.

Similarly, by applying boundary conditions to solve the equations of motion, we derive the condensates as a function of the isospin chemical potential $\mu_I$. The outcomes are presented in Fig. \ref{fig:condensate}. As illustrated in Fig. \ref{subfig:condensate}, when the isospin chemical potential exceeds the mass threshold of the pion meson $m_\pi$, the condensates undergo significant changes. With further increases in the isospin chemical potential, sigma condensation gradually decreases, while pion condensation correspondingly increases. The combination of these two condensates $\Tilde{\sigma}=\sqrt{\sigma^2+\pi^2}$ forms an approximate ``chiral circle'', as intuitively displayed in Fig. \ref{subfig:circle}. Both the SW and HW models predict similar condensation patterns.

It should be noted that the results obtained in this paper differ from the results in Ref. \cite{Cao:2020ske}. In that paper, at high enough isospin densities, the pion meson condensation tends to vanish, showing a similar effect with the baryon density effect on chiral condensate. At the same time, the shape of the formed "chiral circle" also differs. The reason for this discrepancy lies in the different approximations used in this paper compared to \cite{Cao:2020ske}. Specifically, in Ref. \cite{Cao:2020ske}, the gauge field $V$ describing the isospin current is incorporated into the background, resulting in the geometric structure of an AdS-RN black hole. In this paper, however, we follow the method of \cite{Nishihara:2014nsa,Nishihara:2014nva}, solving the probe action with the gauge field $V$ included, while keeping the background as an AdS-Schwarzschild black hole with isospin charge. Of course, one might check the validites of the two approximations by doing a full back-reaction analysis, which is out of the scope of the current work and will be left for the future.

\section{Pionic equation of state}
\label{sec:eos}

As discussed in the previous section, pion condensate emerges in the system when the non-zero isospin chemical potential reaches and exceeds the pion mass threshold. In this section, we explore the equation of state of the system with pion condensate through holographic QCD models. Based on the holographic duality principle, the partition function in gravitational theory is equivalent to quantum field theory on the boundary, that is, $\mathcal{Z}_\text{Gra}=\mathcal{Z}_\text{QCD}$. As a consequence of this equivalence, we can obtain the isospin density $n_I$ and the axial vector condensate, as follows
\begin{eqnarray}\label{eq:nI}
    n_I &=& \int dz \frac{\partial \mathcal{L}_5}{\partial \mu_I} =\int dz \frac{1}{z}(v\Pi^2-a_2\Pi\Sigma), \\
    \langle \sigma \rangle_A &=& \int dz \frac{\partial \mathcal{L}_5}{\partial \mu_A} =\int dz \frac{1}{z}(a_2\Sigma^2-v\Pi\Sigma), \label{eq:sigmaA}
\end{eqnarray}
where we introduce the source term $\mu_A$ for the axial vector field at the AdS boundary. During the calculation of axial-vector condensate, $\mu_A$ is incorporated into the model, and after completing the calculation steps it is reset to 0. 

\begin{figure*}[!tbp]
    \centering
    \includegraphics[width=0.48\textwidth]{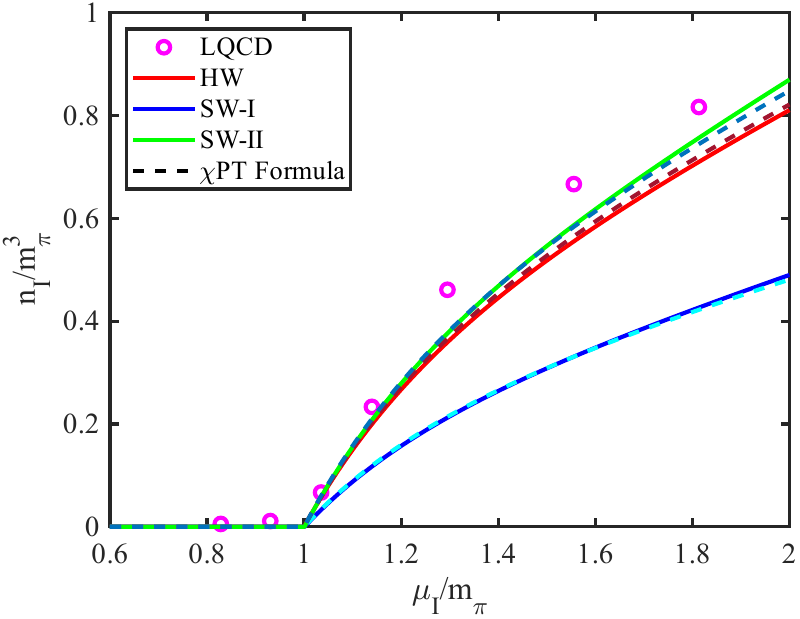}
    \includegraphics[width=0.48\textwidth]{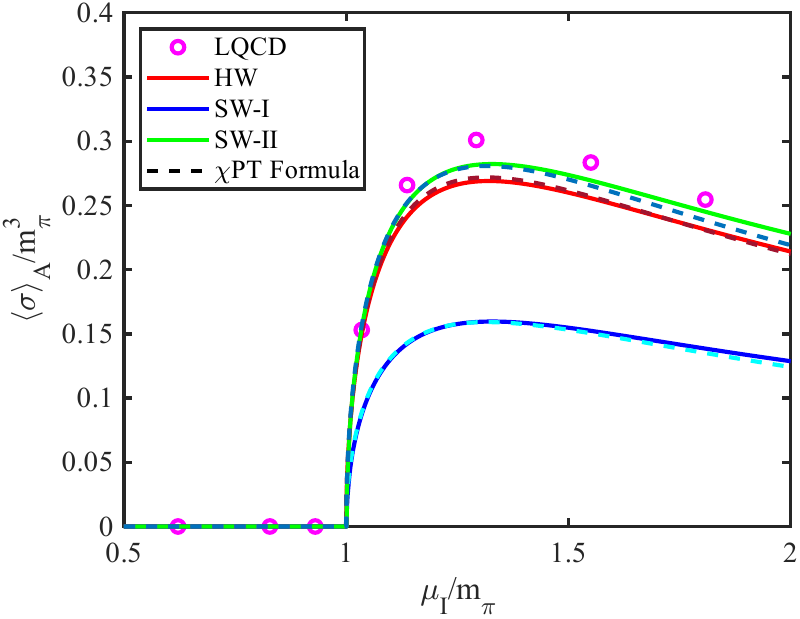}
    \caption{\label{fig:rho} The isospin density and axial vector condensate as a function of isospin chemical potential $\mu_I$ in HW and SW model.
    The black dashed lines are calculations from $\chi$PT formulas \eqref{eq:chipt-1} and \eqref{eq:chipt-2}.}
\end{figure*}

The computational results of the holographic QCD model are presented in Fig. \ref{fig:rho}. The circular markers in the figure represent the lattice QCD data \cite{Brandt:2018bwq}, while the solid lines correspond to the predicted results of HW, SW-I, and SW-II, respectively. The dashed lines in the figure depict the results derived from Chiral Perturbation Theory ($\chi$PT). At zero temperature, according to $\chi$PT \cite{Son:2000xc}, the expressions for isospin density and axial-vector condensation can be simplified to
\begin{eqnarray}
    n_I&=&f_\pi^2\mu_I\bigg(1-\frac{m_\pi^4}{\mu_I^4}\bigg)\Theta(\mu_I-m_\pi), \label{eq:chipt-1}\\
    \langle\sigma\rangle_A&=&\frac{f_\pi^2 m_\pi^2}{\mu_I}\sqrt{1-\frac{m_\pi^4}{\mu_I^4}}\Theta(\mu_I-m_\pi),\label{eq:chipt-2}
\end{eqnarray}
with pion decay constant $f_\pi$ and Heaviside step function $\Theta$. The different dashed lines in the figure correspond to the results obtained from the pion decay constants of different models \cite{Fang:2016nfj,Liang:2023lgs}. 

From Fig. \ref{fig:rho}, it can be observed that the predictions of the HW model and the SW-II model match the data of the lattice QCD quite well, while the results of the SW-I model are approximately half of the lattice data. This discrepancy is because the value of the pion decay constant obtained in the SW-I model is 70.7 MeV, which is significantly lower than the results of lattice simulations. In comparison, the predicted values of the HW model and the SW-II model are closer to the experimentally observation 92.4 MeV. The difference in the pion decay constants leads to a deviation between the model prediction and the data.

\begin{figure}[tbp]
    \centering
    \subfigure[]{\label{subfig:ep}
    \includegraphics[width=0.32\textwidth,height=0.2\textheight]{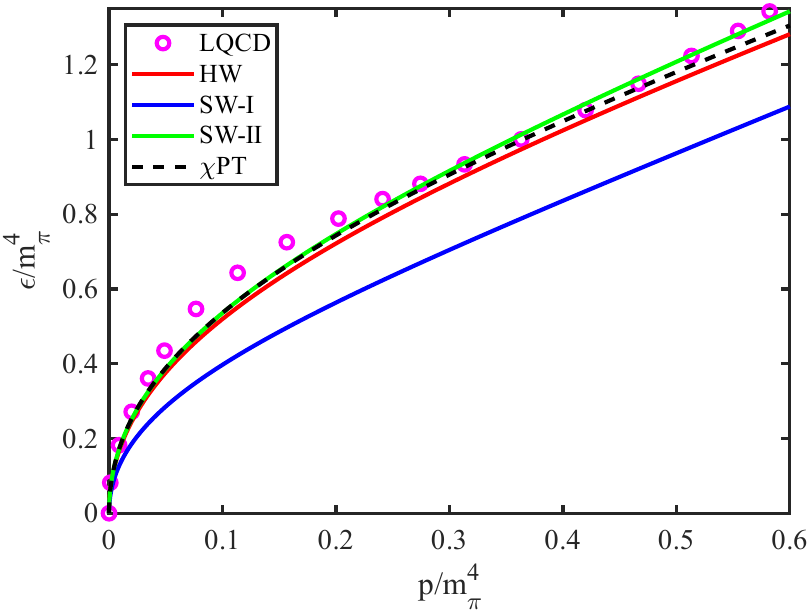}}
    \subfigure[]{\label{subfig:Delta}
    \includegraphics[width=0.32\textwidth,height=0.2\textheight]{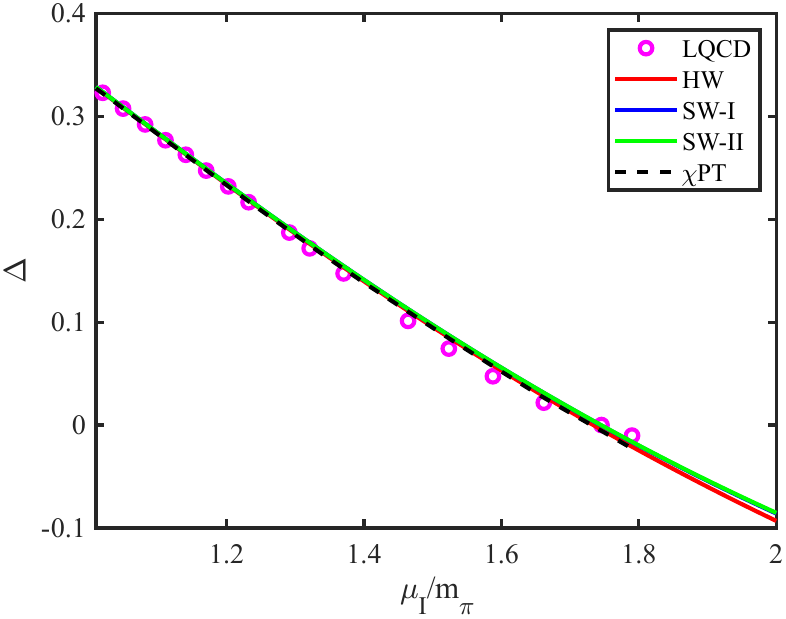}}
    \subfigure[]{\label{subfig:e-3p}
    \includegraphics[width=0.32\textwidth,height=0.2\textheight]{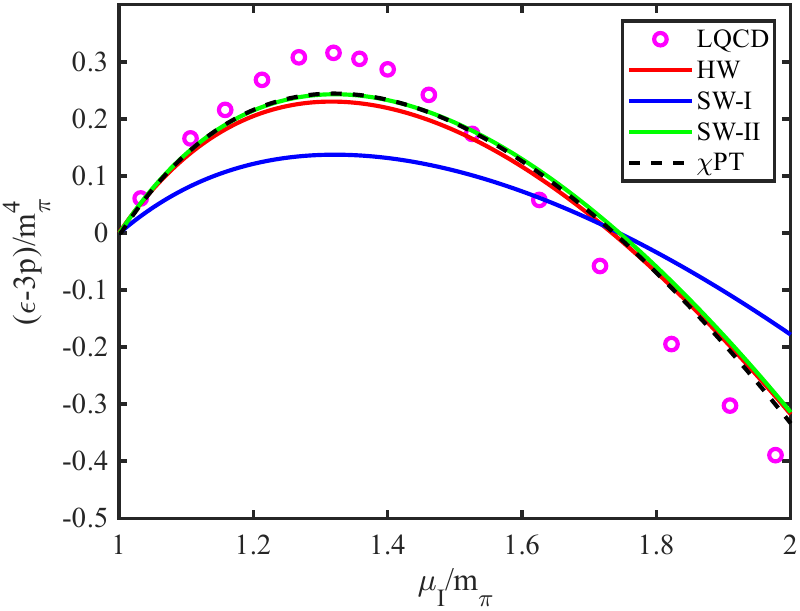}}
    \caption{\label{fig:eos} The equation of state, normalized trace anomalies $\Delta=1/3-p/\epsilon$ and $(\epsilon-3p)/m_\pi^4$ in HW and SW model. The magenta circles represent lattice QCD data \cite{Brandt:2018bwq}. The black dashed lines are calculations from $\chi$PT \cite{Brandt:2022hwy}.}
\end{figure}

At zero temperature, the pionic pressure $p$ and energy density $\epsilon$ can be given by thermodynamic relations
\begin{eqnarray}
    p=\frac{\log \mathcal{Z_\text{QCD}}}{V}=\int_0^{\mu_I} d\mu_I^{\prime} n_I\left(\mu_I^{\prime}\right), \qquad \epsilon=-p+\mu_I n_I.
\end{eqnarray}
The solid lines in Fig. \ref{fig:eos} show the predicted results of our model, while the data from lattice QCD are presented in the circles. Additionally, the Figs. \ref{subfig:Delta} and \ref{subfig:e-3p} display the normalized trace anomalies $\Delta=1/3-p/\epsilon$ and $(\epsilon-3p)/m_\pi^4$, respectively. In these sub-figures, the black dashed lines correspond to the predictions of $\chi$PT \cite{Brandt:2022hwy}. 

It can be observed from Fig. \ref{fig:eos} that the predictions of the HW and the SW-II models are in good agreement with the lattice QCD data \cite{Brandt:2018bwq}, whereas the results of the SW-I model are significantly different from the data, which is mainly due to the difference in the isospin density. For the equation of state, the holographic QCD models' predictions show an approximately linear growth trend at high isospin chemical potential, while the lattice QCD data exhibit different growth characteristics. This discrepancy may arise from the fact that we neglected the contribution of the gravitational background in our calculations and only considered the probe approximation. 

For the trace anomaly $\Delta$, the results of the holographic models are similar to the results of $\chi$PT, both showing a linearly decreasing trend. However, the lattice QCD data show a change in slope when the isospin chemical potential is around $1.7~m_\pi$. For the trace anomaly $(\epsilon-3p)/m_\pi^4$, the predictions of models are closer to the results of the $\chi$PT. However, at the isospin chemical potential of about $1.3~m_\pi$, neither of the models can reach the maximum value shown by the lattice data.

\section{The speed of sound}
\label{sec:cs}

\begin{figure*}[!tbp]
    \centering
    \includegraphics[width=0.48\textwidth,height=0.28\textheight]{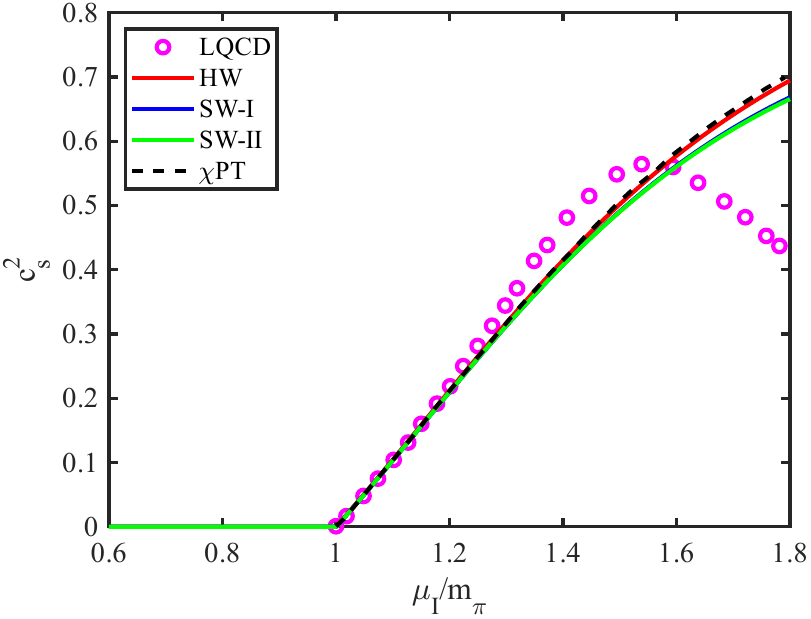}
    \includegraphics[width=0.48\textwidth,height=0.28\textheight]{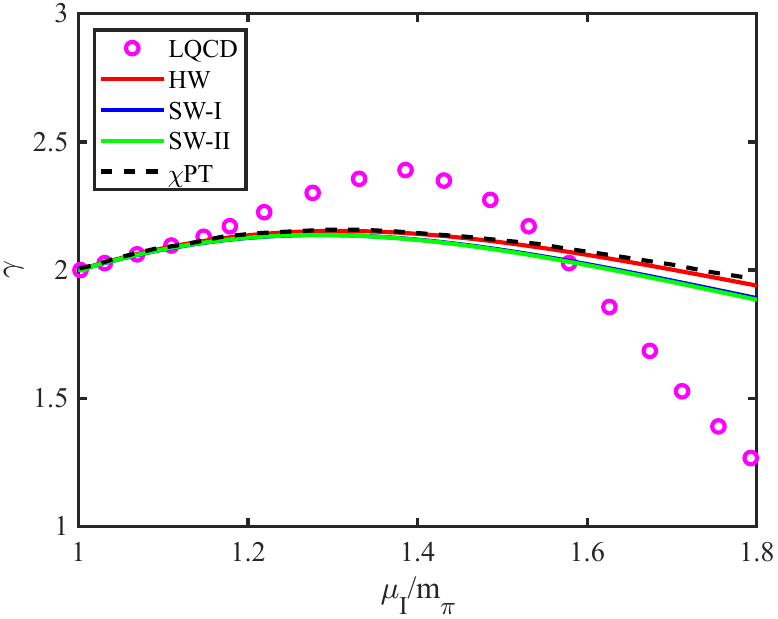}
    \caption{\label{fig:cs} The speed of sound and the adiabatic index as a function of the isospin chemical potential in the HW and SW models. The black dashed lines are calculations from $\chi$PT and the magenta circles are the lattice QCD data from Ref. \cite{Brandt:2022hwy}.}
\end{figure*}

From the equation of state derived in the previous section, we are able to calculate two quantities for astrophysical observations: the speed of sound $c_s$ and the adiabatic index $\gamma$. The square of the speed of sound is defined as
\begin{eqnarray}
    c_s^2=\frac{\partial p}{\partial \epsilon},
\end{eqnarray}
and the adiabatic index is given as
\begin{eqnarray}
    \gamma=\frac{d \log p}{d \log \epsilon}=\frac{\epsilon}{p} c_s^2.
\end{eqnarray}
Fig. \ref{fig:cs} illustrates the results of the speed of sound and the adiabatic index for the isospin chemistry potential and compares them with the lattice QCD data and the $\chi$PT calculations. The results indicate that the predictions of the holographic model are very close to the $\chi$PT calculations, and at low chemical potentials, the holographic model agrees well with the lattice data. However, when the isospin chemical potential exceeds 1.2 $m_\pi$, significant differences begin to emerge between the holographic model and lattice data. Specifically, when the chemical potential reaches 1.5 $m_\pi$, lattice data show that the speed of sound reaches a maximum and then decreases with increasing chemical potential, exhibiting non-monotonic behavior, while the holographic model predicts a continuous increase in the speed of sound. It is worth noting that at two specific $\mu_I$, the normalized trace anomaly is zero, the corresponding sound velocity square is $0$ and $0.65$, both are not equal to 1/3. It needs more investigation whether conformality should be measured by the vanishing trace anomaly \cite{Fujimoto:2022ohj} or $c_s^2=1/3$.

In the conformal limit, the adiabatic index should approach $\gamma=1$, but spontaneous chiral symmetry breaking can cause it to increase to about $\gamma\simeq2$, which is discussed in detail in Ref. \cite{Annala:2019puf}. The adiabatic index value of 1.75 is considered the quark matter bound for distinguishing between two core matter states of neutron stars. The predictions of the holographic model show that the adiabatic index reaches a peak at an isospin chemical potential of about $1.25~m_\pi$, then gradually decreases to approach 2, suggesting that the system may always be near the chiral symmetry broken phase. In contrast, lattice QCD data show an extreme value of the adiabatic index at an isospin chemical potential of about $1.4~m_\pi$, approximately 2.4, and then it drops to 1.3, indicating that the system may have undergone a crossover from BEC to BCS. This difference may be due to the fact that the holographic models do not take into account the contribution of the gluon background in the calculations. If gravitational background were included in the model, similar phase transition behavior might be observed. Furthermore, as mentioned above, within the approximation used in Ref. \cite{Cao:2020ske}, at high isospin density, both the chiral condensate and pion condensate will vanish, which in some sense turns the system from a tightly combined BEC status (meson degree of freedom) to a loosely correlated BCS status (quark degree of freedom). Therefore, similar behavior of crossover from BEC to BCS might appear in that scenario, and in the full back-reaction study as well. However, in this work, we will stick to the current scenario and leave the full back-reaction analysis to the future.

Overall, the HW and SW models provide us with profound insights into strongly coupled pionic matter, but the comparison with lattice QCD data also reveals the limitations of these models. Future research can focus on more precise model parameterization or the development of gravitational frameworks to more comprehensively describe these complex phenomena.

The results from Sec. \ref{sec:eos} and \ref{sec:cs} indicate that both the holographic soft-wall and hard-wall models yield results similar to those of $\chi$PT. In particular, the speed of sound and the adiabatic index from these models align closely and show a consistent discrepancy with lattice data. This suggests a potential equivalence between holographic models and $\chi$PT. As noted in Ref. \cite{Harada:2014lza}, integrating out the heavy Kaluza-Klein modes results in a four-dimensional chiral effective action. A similar discussion is found in Ref. \cite{Albrecht:2010eg}, where the holographic model is used to derive and reproduce the formula \eqref{eq:chipt-1} of $\chi$PT. These findings imply that holographic models share a common ground with $\chi$PT in capturing the physics of strong interactions.

\section{Pion Stars}
\label{sec:star}

In the previous discussion, we studied the pionic equation of state and observed its consistency with lattice QCD data, as well as similar behavior to the results of $\chi$PT. This section will further explore the physical properties of pion stars based on the equation of state.

In the very early universe, asymmetry in lepton flavors led to the phenomenon of pion condensation, which eventually resulted in the formation of pion stars. Pion stars are a type of Bose stars \cite{Wheeler:1955zz,Kaup:1968zz,Jetzer:1991jr}, notable for their existence without the need for new physics beyond the Standard Model. These celestial bodies maintain electrical neutrality and beta equilibrium through the presence of electrons, muons, and their neutrinos. According to Ref. \cite{Stashko:2023gnn}, leptons make a significant contribution to the star's equation of state. However, in this discussion, we will temporarily disregard the impact of leptons on the overall equation of state and focus on other characteristics of pion stars.

\subsection{Mass-radius relation}

The Tolman-Oppenheimer-Volkoff (TOV) equations originate from general relativity, describing the hydrostatic equilibrium state under the assumption of spherical symmetry. Combined with the equation of state, we can calculate the mass $M$ and radius $R$ of pion stars. The form of the TOV equations are as follow,
\begin{eqnarray}
    \frac{d}{d r} p(r)+[p(r)+\epsilon(r)] \frac{G\left[m(r)+4 \pi r^3 P(r)\right]}{r^2\left[1-2 \frac{G m(r)}{r}\right]}&=&0,\\
    \frac{d}{d r} m(r)-4 \pi r^2 \epsilon(r)&=&0,
\end{eqnarray}
with radial coordinate $r$ and 4D Newtonian gravitational constant $G$. The pressure $p$ and the energy density $\epsilon$ both depend on the radial coordinate. Furthermore, $m(r)$ represents the gravitational mass within the radius $r$. The solution of the equations employed the following boundary conditions:
\begin{eqnarray}
    m(r=0)=0, \qquad p(r=r_B)=0,
\end{eqnarray}
with the radius of the star $r_B=R$. The total mass of the pion star is $M=m(r_B)$.

\begin{figure}[tbp]
    \centering
    \subfigure[]{\label{subfig:mr}
    \includegraphics[width=0.48\textwidth,height=0.28\textheight]{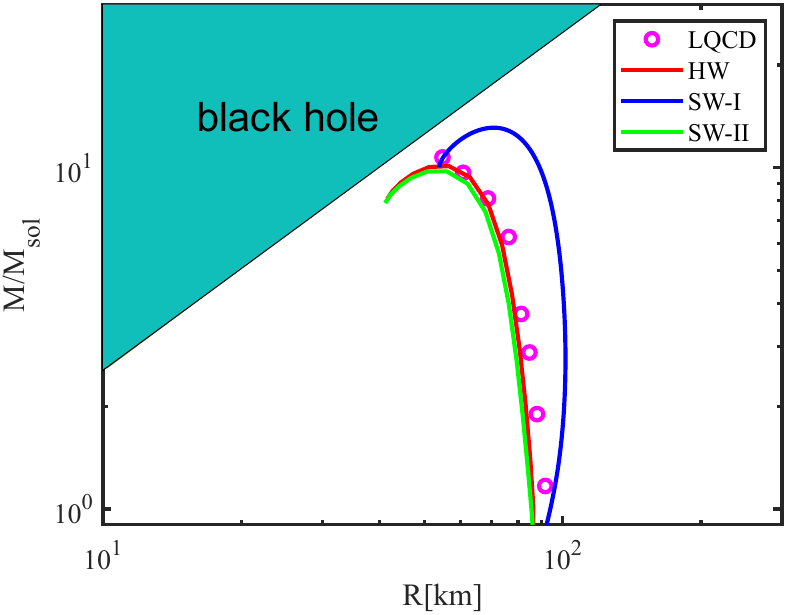}}
    \subfigure[]{\label{subfig:Lambda}
    \includegraphics[width=0.48\textwidth,height=0.28\textheight]{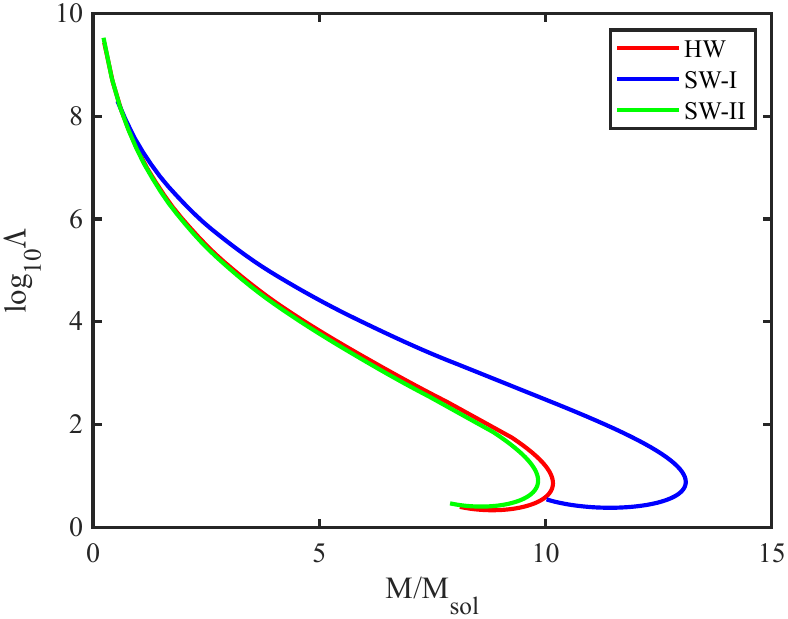}}
    \caption{\label{fig:star} The mass-radius relation and tidal deformability of pion stars. The magenta circles are the mass-radius relation obtained through the equation of state from lattice QCD data Ref. \cite{Brandt:2022hwy}.}
\end{figure}

Using the pionic equation of state obtained in the previous section, we derived the mass-radius relation for pion stars, as shown in Fig. \ref{subfig:mr}. The solid lines of different colors in the figure represent the results of the HW and SW models, while the magenta circles are the mass-radius relations obtained through the equation of state from lattice data. It can be seen from the figure that, due to the proximity of the equation of state to the lattice data, the HW and SW-II models are similar to the lattice results, while the SW-I model predicts a larger upper limit of mass. Overall, the estimated radius of pion stars is between 50-100 kilometers, with the maximum mass reaching 10 times the mass of the sun, which is significantly different from the neutron stars with a radius of 10 kilometers and a maximum mass of 2 times the mass of the sun. If electrons and muons, as well as $\beta$ equilibrium, are further considered, different mass-radius relations for pion stars will be obtained, as detailed in Ref. \cite{Brandt:2018bwq}.

\subsection{Tidal deformability}

Finally, we investigated the dimensionless tidal deformability $\Lambda$, 
\begin{eqnarray}
    \Lambda=\frac{2k_2}{3C^5},
\end{eqnarray}
where
\begin{eqnarray}
    k_2 & = & \frac{8 C^5}{5}(1-2 C)^2[2+2 C(y-1)-y]\left\{2 C[6-3 y+3 C(5 y-8)]\right. \nonumber\\
    & + & 4 C^3\left[13-11 y+C(3 y-2)+2 C^2(1+y)\right]\nonumber\\
    & + & \left.3(1-2 C)^2[2-y+2 C(y-1)] \ln (1-2 C)\right\}^{-1},
\end{eqnarray}
and
\begin{eqnarray}
    C=\frac{GM}{r_B},\qquad y=\frac{r_B\beta(r_B)}{H(r_B)}.
\end{eqnarray}
The functions $H(r)$ and $\beta(r)$ satisfy following equations,
\begin{eqnarray}
    \frac{d}{d r} \beta(r)&-&\frac{2 H(r)}{\left[1-2 \frac{G m(r)}{r}\right]}\left\{-2 \pi G\left[5 \epsilon(r)+ 9 P(r)+\frac{\frac{d}{d r} \epsilon(r)}{\frac{d}{d r} P(r)}[P(r)+\epsilon(r)]\right] \right.\nonumber\\
    &+& \left.\frac{3}{r^2}+\frac{2}{1-2 \frac{G m(r)}{r}} G^2\left[\frac{M}{r^2}+4 \pi r P(r)\right]^2\right\} \nonumber\\
    &-&\frac{2 \beta(r)}{r\left[1-2 \frac{G m(r)}{r}\right]}\left\{-1+\frac{G m(r)}{r}+2 \pi G r^2[\epsilon(r)-P(r)]\right\} =0,
\end{eqnarray}
and
\begin{eqnarray}
    \beta(r)-\frac{d}{d r} H(r)=0,
\end{eqnarray}
with the boundary conditions
\begin{eqnarray}
    H(r \rightarrow 0) \rightarrow a_0 r^2, \qquad \beta(r \rightarrow 0) \rightarrow 2 a_0 r .
\end{eqnarray}
The choice of the constant $a_0$ has no effect on the final result.

The numerical results are shown in Fig. \ref{subfig:Lambda}. It can be seen from the figure that for a mass of 5 solar masses, the tidal deformability is approximately $10^4$.

\section{Conclusion and discussion}
\label{sec:sum}

This study uses holographic hard-wall and soft-wall models to explore the relation between pion condensation, isospin density, axial-vector condensation, and isospin chemical potential. We successfully derived the EoS for pionic matter using thermodynamic relations. Additionally, we calculated the normalized trace anomaly $\Delta$ and $(\epsilon-3p)/m_\pi^4$. Using the EoS, we analyzed the sound speed and adiabatic index, which are crucial for understanding the physical characteristics of pion stars. We also obtained the mass-radius relation and tidal deformability of pion stars, providing quantitative results for studying these bosonic stars.

The results from the holographic models show good agreement with lattice QCD in terms of isospin density, axial-vector condensation, EoS, and normalized trace anomaly. This indicates that holographic models are highly reliable in describing the QCD phase diagram. However, discrepancies were found in the calculations of the sound speed and adiabatic index. At low isospin chemical potentials, the results of the holographic models and lattice QCD coincide. Still, when the isospin chemical potential exceeds $1.5~m_\pi$, the sound speed in lattice data reaches a maximum and then starts to decrease, while it continues to rise in the holographic model. Similarly, when the isospin chemical potential exceeds $1.4~m_\pi$, the adiabatic index in lattice QCD reaches a maximum and then rapidly decreases, while it remains around 2 in the holographic model. These differences might indicate a crossover from BEC to BCS at high isospin chemical potentials, which the holographic model fails to capture.

Further analysis shows that the results from the holographic models are very similar to those from $\chi$PT. In the calculations of isospin density and axial-vector condensation, the results of the holographic models are highly consistent with the approximate formulas of $\chi$PT. In addition, the calculations of the trace anomaly, sound speed, and adiabatic index also align with the predictions of $\chi$PT. Notably, the approximate linear decrease in the trace anomaly, the continuous increase in sound speed with isospin chemical potential, and the adiabatic index remaining around 2 are all consistent with $\chi$PT calculations. This suggests that the holographic model can be considered a five-dimensional $\chi$PT to some extent, offering a new perspective for studying AdS/QCD.

In this study, we employed the holographic approach to the pionic equation of state to calculate the mass-radius relation and tidal deformability of pion stars. Our calculations revealed that the typical radius of a pion star is approximately 100 kilometers, with a mass around $10~M_{\odot}$. The tidal deformability, $\Lambda$, ranges from 10 to $10^8$. The presence of lepton flavor asymmetry in the early universe \cite{Liebling:2012fv,Vovchenko:2020crk} could potentially drive the formation of pion condensation. Additionally, when the energy density in regions of high isospin density reaches a critical threshold, gravitational collapse may occur, leading to the formation of pion stars. However, in this paper, we have not fully explored the formation mechanism or lifetime of pion stars, which are topics that require further investigation.

Our approach differs slightly from the scheme in Ref. \cite{Cao:2020ske}, leading to qualitative differences in the results. Further research is needed to clarify the reasons for these differences and their impact on physical phenomena. The current study neglects the contribution of gluon dynamics, which may be one of the reasons for the inconsistencies between the holographic model and lattice QCD data. Future studies could consider coupling the gravitational background with the system to reproduce lattice QCD results, thereby gaining a more comprehensive understanding of the $T-\mu_I$ phase diagram and properties of strongly interacting pionic matter.

\begin{acknowledgments}
This work is supported by the National Natural Science Foundation of China (NSFC) Grant Nos: 12305136, 12275108, 12235016 , 12221005, the start-up funding of Hangzhou Normal University under Grant No. 4245C50223204075, and the Strategic Priority Research Program of Chinese Academy of Sciences under Grant No XDB34030000, and the Fundamental Research Funds for the Central Universities.
\end{acknowledgments}

\bibliographystyle{unsrt}
\bibliography{ref}

\end{document}